\def\to{\hbox{$\,$--$\,$}}
\def\muspc{\hskip 0.15 em}

\documentstyle[11pt,paspconf]{article}
\input psfig
\begin{document}

\title{The galactic disc age-metallicity relation}
\author{Yuen Keong Ng}
\affil{Padova Observatory, 
Vicolo dell'Osservatorio 5, I-35122 Padua, Italy}

\author{Gianpaolo Bertelli, Giovanni Carraro, Laura Portinari}
\affil{Padova Department of Astronomy, 
Vicolo dell'Osservatorio 5, 
I-35122 Padua, Italy}

\begin{abstract}
New ages are computed for stars in the Solar Neighbourhood from the Edvardsson
et~al.\ (1993) data set. Distances derived from the Hipparcos parallaxes were
adopted to obtain reliable ages (uncertainty less than 12\%) for a subset of
stars. There is no apparent age-metallicity relation for stars with an age less
than 10~Gyr. Only if we consider older stars a slope of
\hbox{$\sim$\muspc0.07~dex/Gyr} appears. This relation is compared with those
obtained from other methods, i.e. galactic open clusters, stellar population
synthesis (star counts), and chemical evolution models. 
\end{abstract}


\section{Introduction}
In the Solar Neighbourhood the metallicity of the stars can be studied in high
detail. Distances need to be known to a 5\% level to get a sub-sample of stars
with reliable ages. In this respect, the age and metallicity 
determination for open
clusters (Carraro et~al.\ 1997) is more reliable, since one is dealing with a
group of stars and the result is less susceptible to individual errors. 
\hfill\break
An age-metallicity relation (AMR) can be obtained from star counts studies,
based on the population synthesis technique (Bertelli et~al.\ 1995, 1996; 
Ng~et~al.\ 1995, 1996, 1997). In such studies all the stars along the line of
sight are considered. The disc is sampled with respect to age and metallicity
in layers with specific effective thickness. In this way indications are
obtained for the disc's chemical evolution. 
\par
\noindent
{\it Our aim is to compare the AMR obtained from various methods and 
to discuss the probable causes for the differences found.} 
\begin{figure}
\setbox1=\vbox{\hsize=7.65cm%
\centerline{\psfig{file=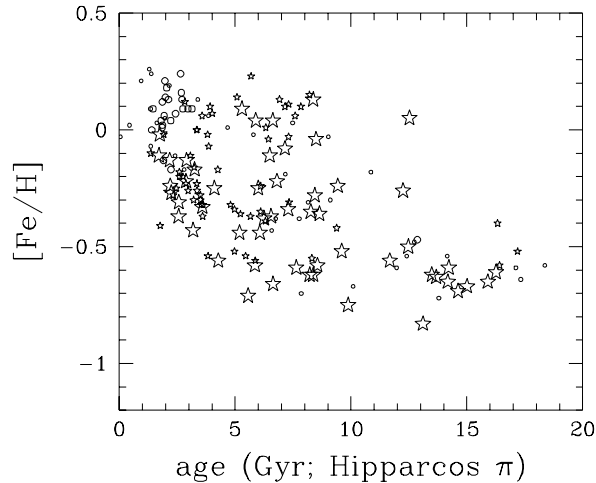,height=6.22cm,width=7.65cm}}
\null\vfill
}
\setbox2=\vbox{\hsize=5.9cm%
\caption{The age-metallicity relation for 
the stars with metallicities from Edvardsson et~al.\ (1993)
and ages by Ng{\muspc\&\muspc}Bertelli (1997).
The open circles denote main sequence stars and 
the asterisks denote sub-giant branch stars. The big
symbols denote the stars with `reliable' ages
for which the uncertainty is less than 12\%.
\null\hfil 
}\vfill}
\centerline{\copy1\quad\copy2}
\end{figure}
\par

\section{Solar Neighbourhood}
Ng{\muspc\&\muspc}Bertelli (1997)
computed new ages for the stars from the Edvardsson et~al.\ (1993)
data set. First we studied the effects due to a change of isochrones.
Then the ages were derived using the distances obtained from
the Hipparcos parallaxes (ESA 1997). 
Figure~1 shows the resulting AMR, which has a slope
of \hbox{$\sim$\muspc0.07~dex/Gyr}.

\section{Open clusters}
An updated 
compilation of open clusters can be found in Carraro et~al.\ (1997). 
The ages for the clusters are all determined 
from fits with the Bertelli et~al.\ (1994) isochrones,
using the synthetic CMD technique.
The main advantage of this compilation is the homogeneity 
of the sample: ages, metallicities and positions
in the galactic plane are all obtained in the same fashion.
This homogeneity is not guaranteed in other, larger samples.
However, it is not possible to gather a complete sample for the old,
open clusters in the galactic disc, because of strong selection effects
mainly related to the past dynamical history  
of the galactic disc: disruption of the clusters.
\par
By means of Multivariate Data Analysis we studied the correlations
among the cluster parameters. 
We considered the four-dimensional parameter space of
age, metallicity, $z$-coordinate and
radial distance R from the galactic centre. 
The results indicate that all four parameters have
a non-negligible weight.
Details of the analysis and their planar projections
are presented in Carraro et~al.\ (1997). 

\section{Stellar population synthesis} 
Synthetic Hertzsprung-Russell diagrams (HRDs) are generated with 
the stellar population synthesis technique.
This is a powerful method in studies of the properties 
of resolved stellar populations.
The so-called HRD galactic software telescope (HRD-GST)
is developed to study the stellar populations in our Galaxy
(Ng et~al.\ 1995). The basis is formed by the latest
evolutionary tracks calculated by the Padova group
(Bertelli et~al.\ 1994 and references cited therein). 
Through a galactic model synthetic 
Colour-Magnitude diagrams are generated.
\par
The primary goal of the HRD-GST is to determine the interstellar
extinction along the line of sight and to obtain constraints on
the galactic structure and on the age-metallicity of the
different stellar populations distinguished in our Galaxy.
The results obtained thus far 
are reported in various papers 
(Bertelli et~al.\ 1995, 1996; Ng et~al.\ 1995\to1997). 

\section{Chemical evolution models}
The AMR for nearby stars is a standard constraint for chemical evolution models
of the Solar neighbourhood. 
Chemical models are aimed at reproducing the enrichment history of
our and other galaxies, giving clues about poorly known processes like star
formation, infall and so forth.
In our Galaxy the resolution on the AMR and on other observational constraints
is high enough
to calibrate model parameters like the star formation rate, the initial mass
function, the infall time-scale. 
With suitable choices of such parameters, all chemical models are basically 
able to reproduce the average AMR
(see Fig.~2, left panel); the major problem is to reproduce the
observed scatter about the average relation (van den Hoek{\muspc\&\muspc}de
Jong 1997). Chemo-dynamical models seem to be a promising improvement. 
\begin{figure}
\setbox1=\vbox{\hsize=6.5cm%
\centerline{\psfig{file=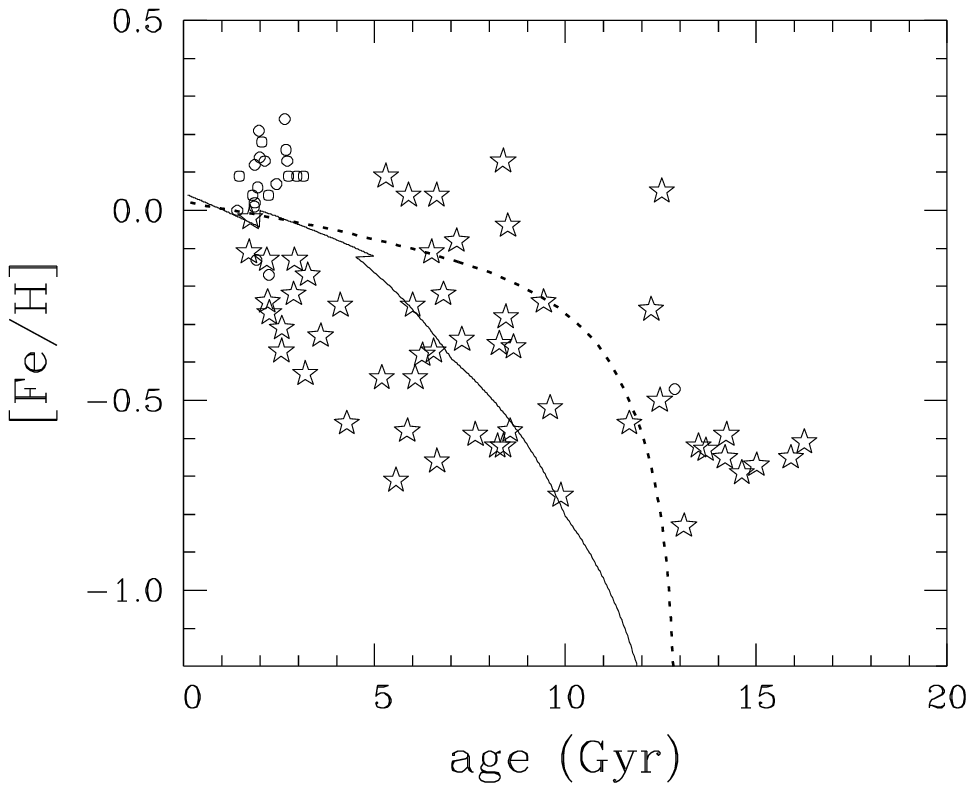,height=5.28cm,width=6.5cm}}
\null\vfill}
\setbox2=\vbox{\hsize=6.5cm%
\centerline{\psfig{file=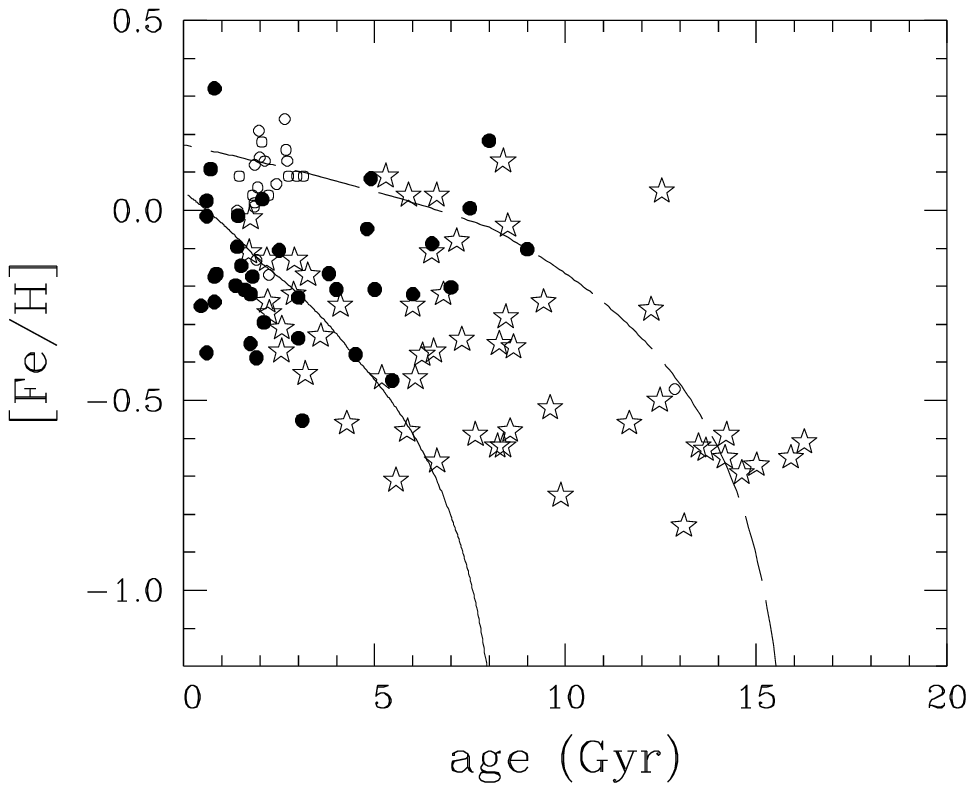,height=5.28cm,width=6.5cm}}
\null\vfill}
\centerline{\copy1\quad\copy2}
\setbox3=\vbox{\hsize=13.7cm%
\caption{{\it Left panel}:
AMR for the 
stars in the Solar Neighbourhood together with 
the prediction of a chemical evolution 
model from Portinari et~al.\ (1997; dotted line), 
and the relation for the galactic disc 
from galactic structure studies
by Ng et~al.\ (1997; solid line).
\hfill\break
{\it Right panel}: AMR for the 
stars in the Solar Neighbourhood together with
the relation for open clusters
and a suggested improvement for the HRD-GST.
The open clusters are corrected for a
radial metallicity gradient of --0.07~dex~kpc$^{-1}$.\null\hfill}
}
\centerline{\copy3}
\end{figure}

\section{On gradients and galactic evolution}
Details about the comparison of the various AMRs can be found in 
Carraro et~al.\ (1997).
Here we only focus on a few points.
A comparison of the AMR obtained from the 
HRD-GST with the one obtained for the Solar Neighbourhood
indicates that the HRD-GST
tends to follow the metal-poorer trend,
which is more populated and is therefore given a larger weight in star counts
(Fig.~2, left panel).
The large spread in metallicity at any particular age 
is likely intrinsic to the disc. It 
cannot be due to the overlap of other galactic components,
because their contribution is negligible.
Figure~2 (right panel) displays two relations, which 
cover respectively the lower and upper metallicity ends of the AMR.
It suggests that the scatter can be studied by means of the HRD-GST
adopting a two-component description.
The first component can be 
associated with the beginning of Galaxy formation,
13\to16~Gyr ago (note that the ages of the oldest
stars might be overestimated, Ng{\muspc\&\muspc}Bertelli, 1997).
The second component was formed 8\to9~Gyr ago and is possibly
induced by the formation of the galactic `bar' (Ng et~al.\ 1996).
\par
Figure~2 (right panel) also displays the AMR of the open clusters
after correction for the 
present day radial gradient.
An unweighted least-squares fit to the clusters
younger than 2~Gyr yields $-0.07$~dex~kpc$^{-1}$; 
within the uncertainties 
the average correction 
is independent of age (Carraro et~al.\ 1997).
We did not apply any correction for the vertical
gradient, since its value or even its existence are 
not clearly established; an apparent vertical gradient
is likely due to insufficient discrimination between age groups.
Figure~2 (right panel) shows that the AMRs of 
open clusters and stars are in good
agreement, both showing a similar trend in the scatter.
Both relations show a lack of scattered points
in the metal-rich side in the age range 3\to5~Gyr and/or
an excess of relatively metal-rich objects in the range 5\to9~Gyr.
This apparent `U-shape' might provide clues about infalling 
and/or merger events.

\vfill
\acknowledgments
The research is supported by 
the Italian Space Agency (ASI), the Italian Ministry of University
and Scientific \& Technological Research (MURST), 
the TMR network of the European Community (grant ERBFMRX-CT96-0086), 
and the Italian National Council of Research (CNR--GNA).

\vfill

\vfill
\begin{references}
\reference Bertelli G., Bressan A., Chiosi C., Fagotto F., Nasi E., 1994,
A\&AS 106, 275
\reference Bertelli G., Bressan A., Chiosi C., Ng Y.K., Ortolani S., 
1995, A\&A 301, 381
\reference Bertelli G., Bressan A., Chiosi C., Ng Y.K., 1996, A\&A 310, 115
\reference Carraro G., Chiosi C., 1994, A\&A 287, 761
\reference Carraro G., Ng Y.K., Portinari L., 1997, MNRAS, {\it submitted}
({\tt astro-ph/9707185})
\reference Edvardsson B., Andersen J., Gustafsson B., et~al., 1993,
A\&A 275, 101
\reference ESA, 1997, The {\it Hipparcos}\/ and {\it Tycho}\/ 
Catalogue, ESA SP-1200
\reference van den Hoek L.B., de Jong T., 1997, A\&A 318, 231
\reference Ng Y.K., Bertelli G., 1997, A\&A {\it in press}
\reference Ng Y.K., Bertelli G., Bressan A., Chiosi C., Lub J., 1995,
A\&A 295, 655
\reference Ng Y.K., Bertelli G., Chiosi C., Bressan A., 1996, A\&A 310, 771
\reference Ng Y.K., Bertelli G., Chiosi C., Bressan A., 1997, A\&A 324, 65
\reference Portinari L., Bressan A., Chiosi C., 1997, in preparation
\end{references}
\end{document}